\documentclass[pr,twocolumn,preprintnumbers,nofootinbib,showpacs,amsmath,amssymb,superscriptaddress]{revtex4-1}
\pdfoutput=1
\usepackage{psfrag,graphicx}
\usepackage[
      colorlinks=true,
      linkcolor=blue,
      urlcolor=blue,
      filecolor=black,
      citecolor=blue,
            ]{hyperref}
\usepackage{color}
\usepackage{amsfonts}
\usepackage{amssymb}
\usepackage{epstopdf}
\usepackage{dsfont}
\usepackage{epsfig}
\newcommand{\Tr}{\text{Tr}}

\begin{document}
\title{Insulator/metal phase transition and colossal magnetoresistance in holographic model}
\author{Rong-Gen Cai}
\email{cairg@itp.ac.cn}
\author{Run-Qiu Yang}
\email{aqiu@itp.ac.cn}
\affiliation{State Key Laboratory of Theoretical Physics,Institute of Theoretical Physics,\\
 Chinese Academy of Sciences,Beijing 100190, China.}

%\pacs{PACS}
%\keywords{keywords}
%\preprint{preprint}
%%%%%%%%%%%%%%%%%%%%%%%%%%%%%%%%%%%%%%
\begin{abstract}
%%%%%%%%%%%%%%%%%%%%%%%%%%%%%%%%%%%%%%
Within massive gravity, we construct a gravity dual for insulator/metal phase transition and colossal magnetoresistance (CMR) effect found in some manganese oxides materials. In heavy graviton limit,  a remarkable magnetic-field-sensitive DC resistivity peak appears  at the Curie temperature, where an insulator/metal phase transition  happens and the magnetoresistance is scaled with the square of field-induced magnetization. We find that metallic and insulating phases coexist below the Curie point and the relation with the electronic phase separation is
discussed.
%%%%%%%%%%%%%%%%%%%%%%%%%%%%%%%%%%%%%%
\end{abstract}
%%%%%%%%%%%%%%%%%%%%%%%%%%%%%%%%%%%%%%
\maketitle
%\tableofcontents

%%%%%%%%%%%%%%%%%%%%%%%%%%%%%%%%%%%%%%
\noindent

\section{Introduction}
In recent years, the holographic correspondence~\cite{Maldacena:1997re,Gubser:1998bc,Witten:1998qj,Witten:1998qj2}, relating  a weak coupling gravitational theory in a (d + 1)-dimensional asymptotically anti-de Sitter (AdS) space-time to a d-dimensional strong coupling conformal field theory (CFT) in the AdS boundary, has been extensively investigated  and some remarkable progresses
have been made in  condensed physics systems~\cite{Gubser:2008px,Hartnoll:2008vx,Lee:2008xf,Liu:2009dm,Cubrovic:2009ye}.  For a recent review on the holographic superconductor/superfluid models, please refer to Ref.~\cite{Cai:2015cya}. Very recently,  the present authors and their collaborators have realized the paramagnetism/ferromagnetism and paramagnetism/antiferromagnetism phase transitions in holographic models by introducing a massive 2-form field in an AdS black brane background and some interesting magnetic properties of the models have been investigated in a series of papers~\cite{Cai:2014oca,Cai:2015bsa,Cai:2014jta, Cai:2014dza,Cai:2015mja,Cai:2015xpa,Cai:2015jta}.  In this Letter, we will provide a new application of the holographic AdS/CFT correspondence by implementing the metal/insulator phase transition and the colossal magnetoresistance (CMR) effect  found in some manganese oxides materials in a holographic model .

Complex magnetic materials showing strong magnetoresistance have simultaneously been the focus of the attentions of the magnetic recording industry and the study of strongly correlated electron systems. Particularly, the study of the manganites such as A$_{1-x}$B$_x$MnO$_3$ (A= La, Pr, Sm, etc. and B = Ca, Sr, Ba, Pb), which exhibit the ``colossal'' magnetoresistance effect, is among the main areas of research in strongly correlated electron systems~\cite{Urushibara,Uehara,MBS,Dagottoa,Nagaev,Cengiz,Mukherjee}. These materials show remarkable magnetoresistivity and an insulator (or semiconductor)/metal phase transition associated with a paramagnetic/ferromagnetic phase transition, which has a completely different physical origin from the ``giant" magnetoresistance observed in layered and clustered compounds. These materials are currently being  intensively investigated by a sizable fraction of the condensed matter community, and its popularity is reaching the level comparable to  the high-temperature superconducting cuprates.

After great efforts in recent years, mainly  through computational and mean field studies for realistic models, considerable progress has been achieved in understanding the curious properties of those compounds. However, a fully quantitative understanding of the CMR effect is still a challenge, much work remains to be carried out and  it is the subject of current active research~\cite{Dagottoa2}. The holographic duality provides an alternative method for this type of strong correlated phenomena.  In this paper, we will make a first attempt to build a holographic model to understand the CMR effect.

%%%%%%%%%%%%%%%%%%%%%%%%%%%%%%%%%%%%%%%%%%%%%%%%
\section{Holographic model}
Before presenting our holographic model, let us  make a brief analysis about how to build a holographic description for such a phenomenon. Firstly, before ferromagnetic phase transition happens, CMR materials are in insulating phase where DC resistivity is finite and increases with decreasing the temperature. So the translation symmetry is broken otherwise no scattering happens and DC resistivity is divergent. More important  is that, more and more results from experiments show that the insulating phase in CMR is charge ordered, in which charges are localized and form inhomogeneous structures~\cite{Dagottoa}. To realize the inhomogeneity  for the CMR effect is a  challenge both in condensed matter theory and holographic description. Fortunately, just as pointed out recently  in Ref.~\cite{Blake:2013owa},  momentum relaxation by breaking the translation invariance can be achieved by introducing a mass term of graviton in the bulk so that the macroscopic DC resistance becomes
finite. This provides us with a very simple holographic model  to study  macroscopic DC resistance in some inhomogeneous materials without involving some complicated computations. Secondly, in general, only breaking the translational invariance cannot  lead to an insulating resistivity. Meffort and Horowitz~\cite{Mefford:2014gia} proposed a  simple framework to have the insulating behavior in general relavity without breaking translational invariance, where a real scalar field is coupled to an U(1) gauge field. But Meffort-Horowitz  model is only valid in the case of zero charge density. Thus a natural choice
to build a holographic insulator model with finite charge density is to consider the Meffort-Horowitz model in a massive gravity theory. Finally, motivated by our previous work about DC resistivity in the paramagnetism/ferromagnetism phase transition in the probe limit~\cite{Cai:2015jta}, the model with massive 2-form field coupled with Maxwell field shows metallic ferromagnetic phase in low temperatures. This provides us with a mechanism to describe metallic resistivity after the ferromagnetic phase transition.

Based on these considerations, we present the model with the massive 2-form field-Maxwell-dilaton theory in a massive gravity with a negative cosmological constant. The action can be written as,
\begin{equation}\label{insferr}
\begin{split}
&S=\frac1{16\pi G}\int d^4x\sqrt{-g}[\mathcal{R}+\frac{6}{L^2}-(L_{\text{ins}}+\lambda^2L_{\text{ferr}})+L_{\text{mg}}],\\
&L_{\text{ins}}=e^{-2g_0\psi}F_{\mu\nu}F^{\mu\nu}+\frac12(\nabla\psi)^2+\frac{m_2^2}2\psi^2,\\
&L_{\text{ferr}}=\frac{(dM)^2}{12}+\frac{m_1^2}4M_{\mu\nu}M^{\mu\nu}+\frac12M^{\mu\nu}F_{\mu\nu}+\frac J8V(M),\\
&L_{\text{mg}}=\alpha\Tr\mathcal{K}+\beta[(\Tr\mathcal{K})^2-\Tr\mathcal{K}^2].
\end{split}
\end{equation}
Here $L$ is the AdS radius and $G$ is the Newtonian gravitational constant. Without loss of generality, we can set $L=1/16\pi G=1$. $\lambda, J, \alpha, \beta$ and $g_0$ are all model parameters. $\mathcal{R}$ is scalar curvature, $A_\mu$ is the U(1) gauge field with field strength $F_{\mu\nu}=2\nabla_{[\mu}A_{\nu]}$.  $\psi$ is a dilaton field with squared mass $m_2^2$. $M_{\mu\nu}$ is a 2-form field with squared mass $m_1^2$ and a nonlinear potential $V(M)$. $dM$ is the exterior differential of 2-form field $M_{\mu\nu}$ and $(dM)^2=9\nabla_{[\mu}M_{\nu\tau]}\nabla^{\mu}M^{\nu\tau}$.$L_{\rm mg}$ is the mass term of graviton, where matrix $\mathcal{K}$ is defined in terms of the dynamical metric $g_{\mu\nu}$ and a reference background metric $\mathfrak{f}_{\mu\nu}$ by
\begin{equation}\label{matrixK}
    {\mathcal{K}^\mu}_\nu{\mathcal{K}^\nu}_\tau=g^{\mu\nu}\mathfrak{f}_{\nu\tau}
\end{equation}
This is a special case of the formulation of the massive gravity theory~\cite{deRham:2010ik,deRham:2010kj} presented in~\cite{Hassan:2011vm}.  As shown in Refs.~\cite{Blake:2013bqa,Blake:2013owa}, there is a position-dependent mass of the gravitational perturbations,
\begin{equation}\label{massgravity}
    m^2_g(r)=-2\beta-\alpha /r.
\end{equation}
Ref.~\cite{Blake:2013owa} shows that, in the leading level of lattice,  the graviton mass term is proportional to the square of modulating amplitude multiplied by wave vector in the dual boundary lattice. This gives a physical meaning of graviton mass in holographic model, i.e., $m_g^2$ describes the strength of inhomogeneity  in the dual boundary theory. This consequence plays a key role in our holographic model. In this paper, we will pay attention to two limits, i.e., weak inhomogeneous and strong inhomogeneous cases, which correspond to the cases of $m^2_g\ll1$ and $m_g^2\gg1$, respectively. These two cases admit us to obtain DC resistivity in a simple manner. We will see later, only the case with large value of $m^2_g$ can give rise to the typical CMR effect, which agrees with the fact that there is a strong inhomogeneity with CMR effect in the manganite.

The parameter $\lambda$ can be understood as the coupling strength between the polarization field $M_{\mu\nu}$ and background Maxwell field strength. In the effective  action of string tehory, the value of the dilaton coupling parameter is taken to be $g_0=1$. The case with  $g_0=\sqrt{3}$ corresponds to the 4-dimensional action by dimensionally reducing from the 5-dimensional Kaluza-Klein theory~\cite{Wiltshire:1994de}. Nonetheless, it is  helpful to set $g_0$ as an arbitrary positive constant here so that we can see what  the role  of dilaton field plays in the model. In addition, the choice of nonlinear potential $V(M)$ is not unique. What we need is that there is a critical temperature, below which the xy-component of $M_{\mu\nu}$ can condense. In this paper, we take the potential as follows,
\begin{equation}\label{poten}
V(M)=({^*}M_{\mu\nu}M^{\mu\nu})^2=[{^*}(M\wedge M)]^2.
\end{equation}
Here $^*$ is the Hodge-star operator. We choose this form just for simplicity. To be stable for both the bulk and boundary theories when $\psi=0$, we require $m^2_g\geq0$ for all $r$~\cite{Davison:2013jba,Vegh:2013sk}. Following~\cite{Vegh:2013sk}, we take the degenerate reference background with $\mathfrak{f}_{xx}=\mathfrak{f}_{yy}=1$ and other components are vanishing. Although the reference background is degenerate,  Ref.~\cite{Vegh:2013sk}  showed that this two-parameters  massive theory is ghost free for the case with $\beta$ mass term, but it has not yet been proven for the  case with $\alpha$ mass term. Therefore in this paper, we will take $\alpha=0$ in order to avoid possible problems with causality. Note that this reference background leads to that there is a conserved energy but no conserved momentum current in the boundary theory.

%%%%%%%%%%%%%%%%%%%%%%%%%%%%%%%%%%%%%%%%%%%%%%%%%%%%%%%%%%%%%%%%
\section{Background equations and perturbations}
Now we are in the position to calculate the DC resistivity from our holographic model~\eqref{insferr}. For this, we assume the dynamic metric has following form,
\begin{equation}\label{geo1}
\begin{split}
ds^2&=-r^2fe^{-\chi}dt^2+\frac{dr^2}{r^2f}+r^2(dx^2+dy^2),\\
\end{split}
\end{equation}
where $f$ and $\chi$ are two functions of $r$. Suppose the solution has a horizon at $r_h$,  the associated  temperature then is $T=r_h^2f'e^{-\chi/2}/4\pi$. We take the following self-consistent ansatz  for matter fields,
\begin{equation}\label{ansatz1}
\begin{split}
&A_\mu=\phi(r)dt+Bxdy,~\psi=\psi(r)\\
&M_{\mu\nu}=-p(r)dt\wedge dr+\rho(r)dx\wedge dy.
\end{split}
\end{equation}
Here $B$ is a constant magnetic field and it will be viewed as the external magnetic field in the boundary field theory. Put the ansatz and the dynamic matric in~\eqref{geo1} into the action~\eqref{insferr}, we can get a set of ordinary differential equations. Near the boundary $r\rightarrow\infty$, the equations give the following asymptotic solutions for matter fields,
\begin{equation}\label{asy0}
\begin{split}
&\rho=\rho_+(\frac{r}{r_h})^{(1+\delta_1)/2}+\rho_-(\frac{r}{r_h})^{(1-\delta_1)/2}+\cdots+\frac{B}{m_1^2},\\
&p=\frac{\sigma r_h^2}{m_1^2r^2}+\cdots,~\phi=\mu-\frac{\sigma r_h}r+\cdots,\\
&\psi=\psi_+(\frac{r}{r_h})^{(\delta_2-3)/2}+\psi_-(\frac{r}{r_h})^{-(\delta_2+3)/2}+\cdots,
\end{split}
\end{equation}
where $\delta_1=\sqrt{1+4m_1^2}$ and $\delta_2=\sqrt{9+4m_2^2}$,  $\mu$ is the chemical potentail, $\sigma$ is the charge density and $\rho_{\pm}$ and $\psi_{\pm}$
are all constants. We impose the regular conditions at the horizon and Dirichlet and source free conditions for matter fields at the boundary of $r\rightarrow\infty$, i.e., $\phi=\mu,\psi_+r_h^{(3-\delta_2)/2}=\Delta$ and $\rho_+=0$.
%
%\begin{equation}\label{bound1}
%\phi=\widetilde{\mu},~\rho_+=0,~~\psi_+=0.
%\end{equation}
%
Without loss of generality, we can set $\mu=1$. Note that nontrivial solution for $\psi$ always exists when $g_0$ and $F_{\mu\nu}$ are both nonzero.

Following Ref.~\cite{Cai:2015jta}, we need that $J<0$ so that $\rho$ can spontaneously condense below a critical temperature when $B=0$. We also need that $\delta_1>1$ and $\delta_2<3$, otherwise, the nonlinear terms of $\rho$ and $\psi$ will play more and more important roles when $r\rightarrow\infty$, which will break the asymptotic AdS$_4$ geometry of space-time and lead to the instability of the dual theory in the UV regiem.

%%%%%%%%%%%%%%%%%%%%%%%%%%%%%%%%%%%%%%%%%%%%%%%
Now let us study how DC conductivity is influenced by temperature and external magnetic field in this model. Because of the planar symmetry in the boundary, the conductivity is isotropic. By the  AdS/CFT
dictionray,  we need only turn on a perturbation of gauge field such as $\delta A_x$ along the $x$-direction with ingoing boundary condition at the horizon.

In the case of weak inhomogeneity $m_g^2\ll1$, the main part of DC resistivity is very simple. In that case, the graviton mass is very small,  gravitation fluctuations suffer from smaller scattering than others. So the DC conductivity is dominated by the background geometry. In other words, we can neglect the fluctuations of matter fields when we compute the DC resistivity. Then following Refs.~\cite{Blake:2013bqa,Blake:2013owa}, we can find that  DC resistivity can be expressed as
\begin{equation}\label{Rweak}
    R_{\text{light}}=m_g^2/[\mu^2(1-\lambda^2/4m_1^2)]+\mathcal{O}(m_g^4).
\end{equation}

The more interesting case is the strong inhomogeneous limit, i.e., $m_g^2\gg1$, which is the case we are really interested in this paper. In such a limit, graviton has very heavy mass so that it is in fact very hard to be excited by fluctuations. Heavy mass term suppress the fluctuation of gravity so that we can fix the background geometry. In this case, the main part of DC resistivity can be obtained by just considering the fluctuations of matter fields. In the DC limit of $\omega\rightarrow0$, we need only consider three perturbations $\delta A_x=\epsilon a_x(r)e^{-i\omega t}$ and $M_{ij}=\epsilon C_{ij}(r)e^{-i\omega t}$ with ingoing conditions at the horizon, where $(i,j)=(r,x), (t,y)$. At the linear order of $\epsilon$, we have following equations in the low frequency limit,
\begin{subequations}\label{pert2}
\begin{align}
C_{ty}''+\frac12\chi'C_{ty}'-\frac{m_1^2C_{ty}}{r^2f}-\frac{4Jp\rho C_{rx}}{r^2}+O(\omega)&=0,\label{pert2a}\\
m_1^2C_{rx}-a_x'-\frac{4Je^\chi p\rho C_{ty}}{r^4f}+O(\omega)&=0,\label{pert2b}\\
[r^2fe^{-\chi/2}(e^{-2g_0\psi}a_x'-\lambda^2C_{rx}/4)]'+O(\omega)&=0,\label{pert2c}
\end{align}
\end{subequations}
where  a prime stands for the derivative with respect to $r$, and $\psi, p$ and $\rho$ determined by the background equations, $O(\omega)$ is the terms with order of $\omega$, and the other equations of gravity parts and matter parts are order of $O(\omega)$ , all of which can be neglected when $\omega\rightarrow0$.

At the boundary $r\to \infty$ with $p, \rho\rightarrow0$, we have the following asymptotic solution for $a_x$,
\begin{equation}\label{pert3}
a_x=a_{x+}+\frac{a_{x-}}{r}+\cdots.
\end{equation}
Following the AdS/CFT dictionary,  we can obtain that electric current is $\langle J\rangle=a_{x-}$ and the DC  resistivity is given by $R=\lim_{\omega\rightarrow0}i\omega a_{x+}/\langle J\rangle$.

In fact,  with the ``membrane paradigm" of black hole, we can directly obtain the DC conductivity from  Eqs.~\eqref{pert2} using the method proposed by Iqbal and Liu in Ref.~\cite{Iqbal:2008by}. To see this, we first note that,
\begin{equation}\label{pert4}
\lim_{r\rightarrow\infty}r^2 f(a_x'-\lambda^2C_{rx}/4)=-(1-\lambda^2/4m_1^2)\langle J\rangle.
\end{equation}
Eq.~\eqref{pert2c} shows that this quantity is conserved along the direction $r$ when $\omega\rightarrow0$. At the horizon, using Eqs.~\eqref{pert2} and the fact that $C_{ty}$ is regular at the horizon when $\omega\rightarrow0$, we have,
\begin{equation}\label{pertb2}
%\begin{split}
\langle J\rangle=\frac{i\omega a_x(r_h)}{1-\frac{\lambda^2}{4m_1^2}}\left[e^{-2g_0\psi_0}-\frac{\lambda^2m_1^2}{4(m_1^4+16J^2e^{\chi_0}p_0^2\rho_0^2/r_h^4)}\right].
%\end{split}
\end{equation}
Here $\chi_0, \psi_0, p_0$ and $\rho_0$ are the initial values of $\psi, p$ and $\rho$ at the horizon, respectively. In the low frequency limit, Eq.~\eqref{pert2c} implies that the electric field is constant, i.e., $\lim_{r=r_h}a_x(r)=a_{x+}$. So we obtain the DC resistivity in heavy graviton limit as,
\begin{equation}\label{DCexp}
\begin{split}
%\frac1{R}=\frac{1}{1-\frac{\lambda^2}{4m_1^2}}\left[e^{-2g_0\psi_0}-\frac{\lambda^2m_1^2}{4(m_1^4+16J^2p_0^2\rho_0^2/r_h^4)}\right].
\frac1{R_{\text{heavy}}}=&(1-\frac{\lambda^2}{4m_1^2})^{-1}\left[e^{-2g_0\psi_0}-\right.\\
&\left.\frac{\lambda^2m_1^2}{4(m_1^4+16J^2e^{\chi_0}p_0^2\rho_0^2/r_h^4)}\right]+\mathcal{O}(1/m_g^2),
\end{split}
\end{equation}
As a self-consistent check, we can take $\lambda=0$. In that case there are only dilaton and Maxwell field. Then we have $R_{\text{heavy}}^{-1}=e^{-2g_0\psi_0}+\mathcal{O}(1/m_g^2)$, which agrees with the exact result $R^{-1}=Z(\psi_0)+\mu^2/m_g^2$ given in  Ref.~\cite{Blake:2013bqa} with daliton coupling $Z(\psi_0)=e^{-2g_0\psi_0}$. From the expression for DC resistivity, we see that when $R_{\text{heavy}}\sim m_g^2$, the heavy graviton limit is broken. In that case, the fluctuations of metric have to be taken into account.

%%%%%%%%%%%%%%%%%%%%%%%%%%%%%%%%%%%%%%%%%%%%%%%
\section{Metal/insulator phase transition and magnetoresistance in strong inhomogeneity }

\begin{figure}[h!]
\begin{center}
\includegraphics[width=0.32\textwidth]{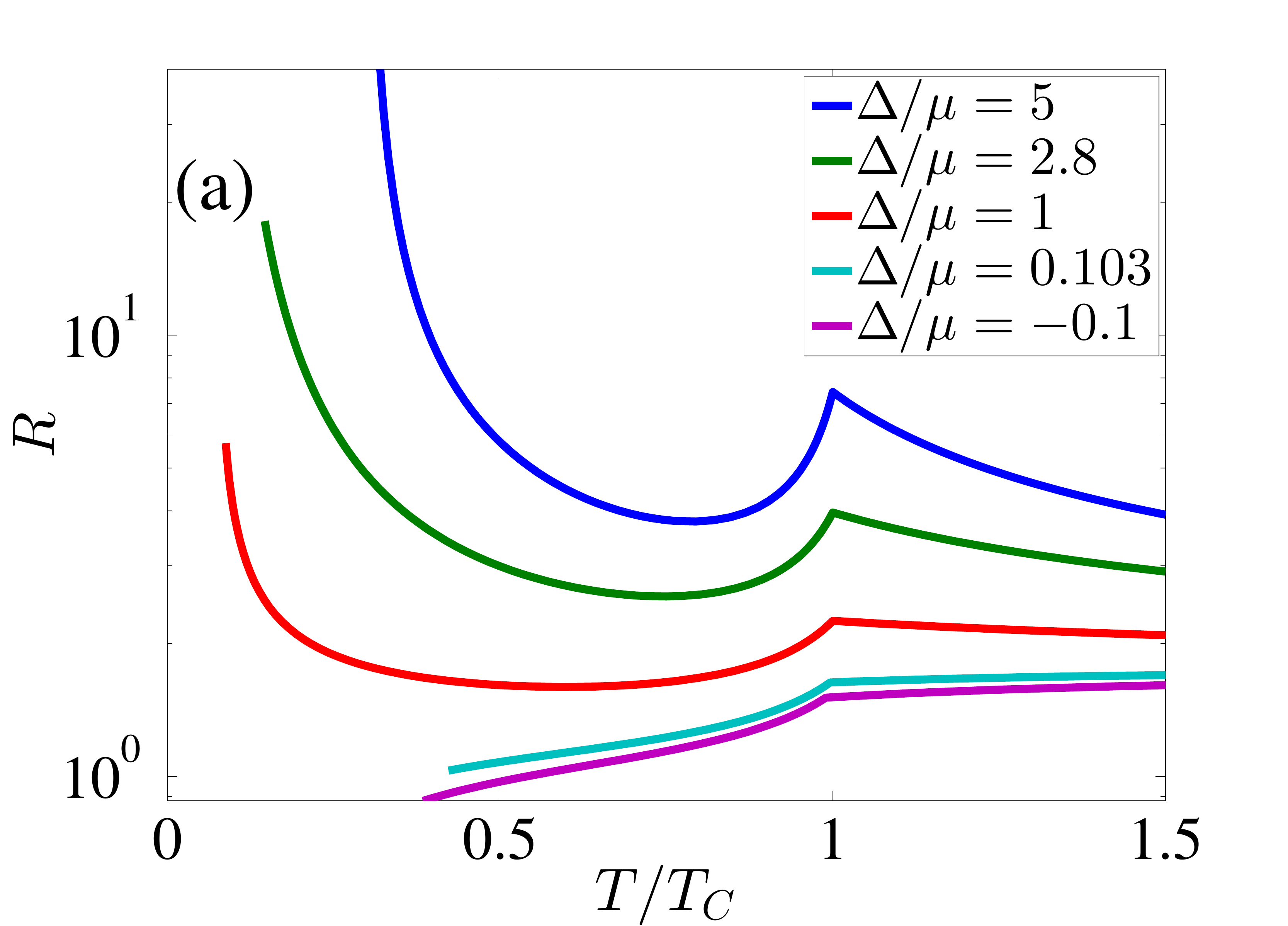}
\includegraphics[width=0.32\textwidth]{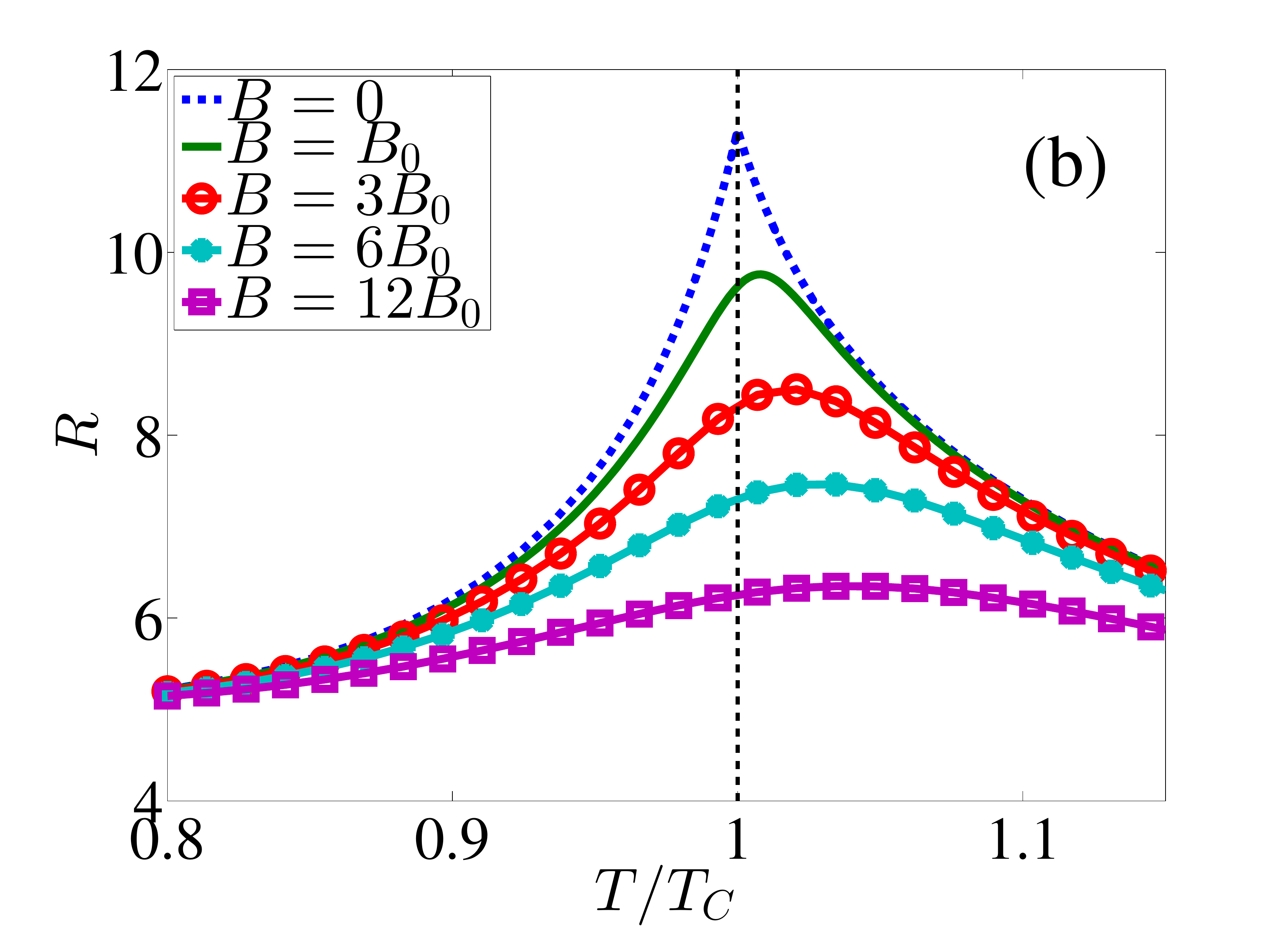}
\includegraphics[width=0.25\textwidth]{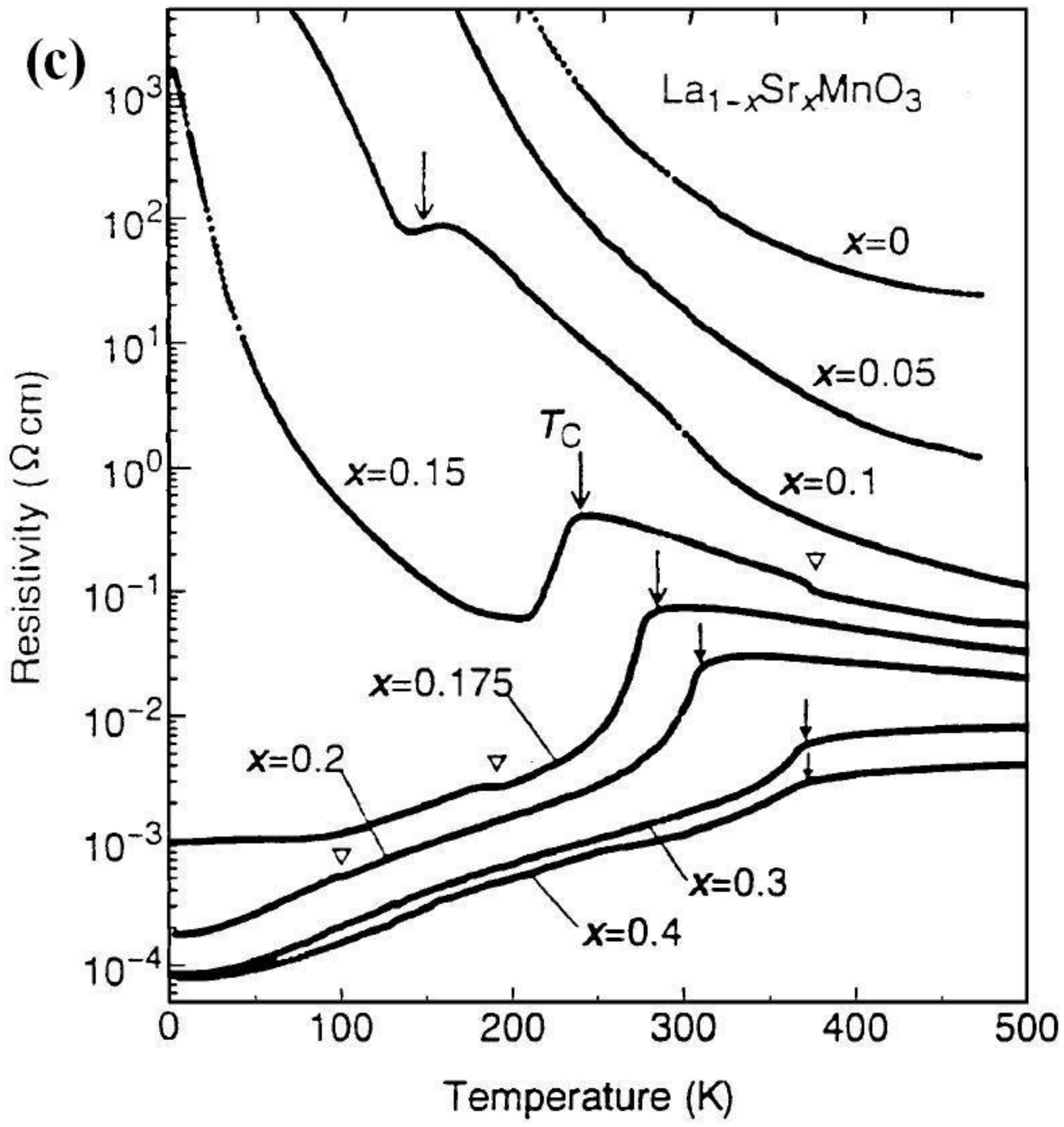}
\includegraphics[width=0.32\textwidth]{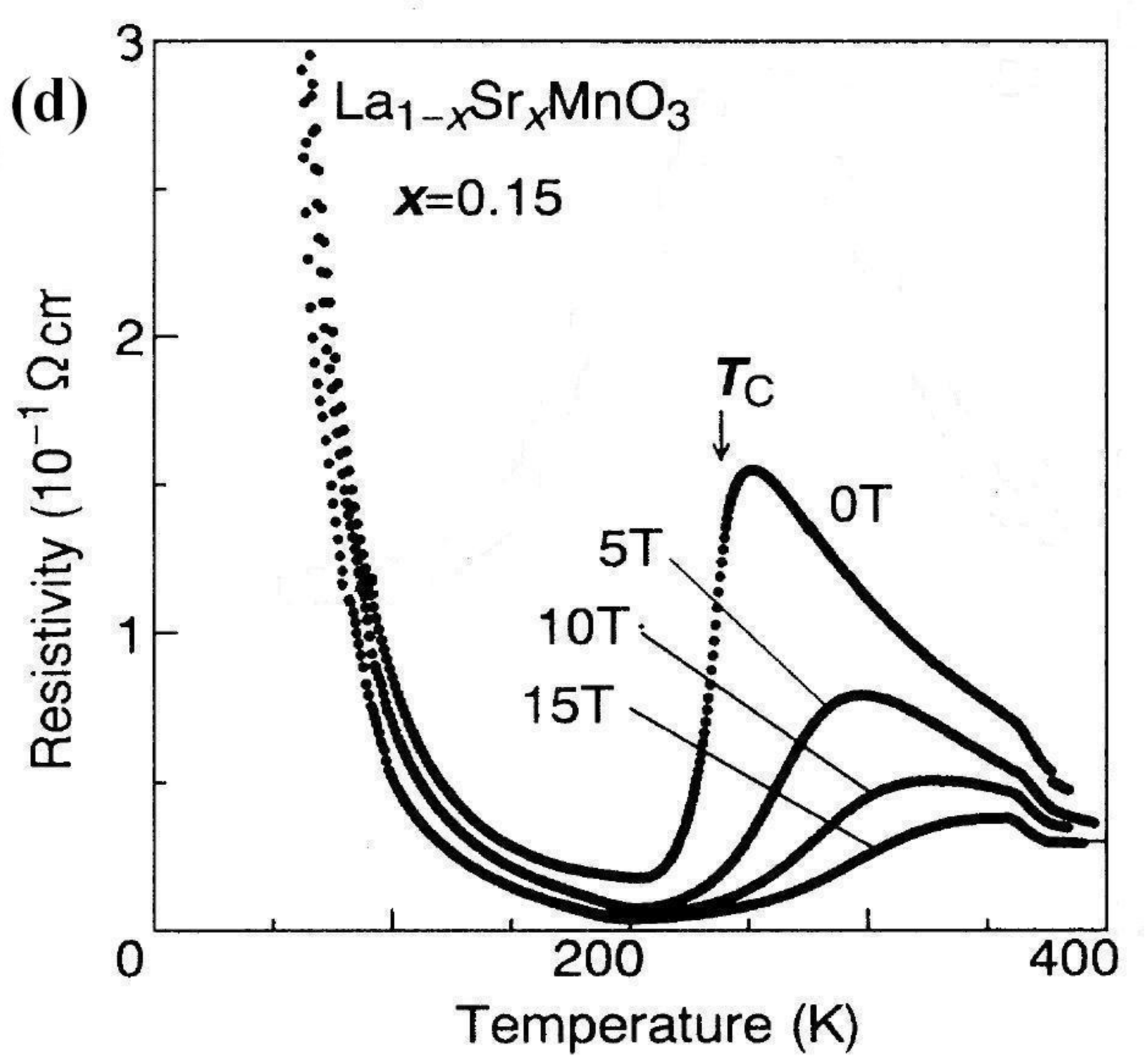}
\caption{(a): The behavior of the DC resistivity vs temperature in different $\Delta$ with $B=0$ and $J=-2$. (b): The DC resistivity vs temperature in different external magnetic field. Here $B_0/T_C^2\simeq5.5\times10^{-4}$, and $J=-1/3$. (c) and (d):  The DC resistivity for La$_{1-x}$Sr$_x$MnO$_3$ as a function of temperature for different doping $x$ and magnetic field. The experimental data are from Ref.~\cite{Urushibara}. }
\label{insul3}
\end{center}
\end{figure}

The physical phase in different temperature depends on the model parameters. As a typical case , we fix $m_1^2=1/3, m_2^2=-2, m_g^2=40, g_0=1$ and $\lambda=3/4$ to compute the DC resistivity at different temperature and small external magnetic field numerically. We first solve the equations of motion from the action and then calculate the DC resistivity numerically.
All the results are shown in  Fig.~\ref{insul3}.

In Fig.~\ref{insul3}(a), we plot  the DC resistivity at zero magnetic field with different dilaton source $\Delta$. There is a critical $\Delta_c\simeq0.103$ (we scan $\Delta$ from 5 to -1, numerical precision restricts our ability to check wider region). When $\Delta>\Delta_c$, there is an insulator/metal phase transition. The resistivity shows an insulator's behavior described by dilaton field when $T>T_C$. When the temperature is lowered to the Curie temperature $T_C$, $\rho$ begins to condense spontaneously and a ferromagnetic phase transition happens. Below and near $T_C$, the resistivity decreases when temperature is lowered, which shows a metal's behavior. Though the behavior of DC resistivity is transformed into  metallic from insulating, the insulating phase described by dilaton field coexists with ferromagnetic metallic phase in the sample. Numerical results show that two different electronic phases can coexist below the Curie temperature. There is a distinct peak at the temperature where spontaneous magnetization begins to appear and an insulator/metal phase transition happens there.  This is just one of characteristic properties of CMR materials in manganese oxides. When $\Delta<\Delta_c$, DC resistivity shows a metallic behavior in the whole temperature (up to the region that numerical computations can be done), though there is still a saltation at $T_C$. What's more, when a small magnetic field $B$ is turned on, we find that the resistivity is very sensitive to the external magnetic field (see Fig.~\ref{insul3}(b)).

Here we emphasize that the heavy graviton limit plays a very important role. Just as mentioned above, for the light graviton case, the behavior of DC resistivity is dominated by fluctuation of graviton, then no such metal/insulator phase transition or CMR effects appear. This is agreement with the fact that  the CMR effect is  due to the fact that two electronic phases mix with each other and form inhomogeneity in nm scale~\cite{Dagottoa,Uehara,SW.Cheonga}.

It is interesting to compare our holographic results with some experimental data of  CMR materials. The resistivity of a typical CMR material La$_{1-x}$Sr$_x$MnO$_3$ is shown in  Fig.~\ref{insul3}(c) and (d). We see that our holographic model gives qualitatively similar results when $x\geq0.1$, which is a powerful evidence to support that this holographic model is a suitable one to describe the CMR effect. Furthermore, in La$_{1-x}$Sr$_x$MnO$_3$,  when $x>x_c\simeq0.2$, the peak of DC resistivity disappears, which is very similar to the case when daliton source $\Delta<\Delta_c$. So this similarity gives us an explanation for $\Delta$ at the dual boundary field, i.e., dilaton field describes the effects of doping.  In addition, we can find from  Eq.~\eqref{DCexp} the following scaling relation for magnetoresistance (MR) in the case of weak magnetic field and $T\rightarrow T_C^+$,
\begin{equation}\label{scal1}
\text{MR}(B)=1-R(B)/R(B=0)\propto \rho_0^2\propto B^2.
\end{equation}
Note that in the region of $T>T_C$, the system is in the paramagnetic phase.  Therefore the magnetic moment $N$ is proportional to $B$ in the weak field case. Then  Eq.~\eqref{scal1} tells us that $\text{MR}(B)\propto N^2$. This result is in complete agreement with the experimental data of  CMR materials~\cite{Urushibara}.

%%%%%%%%%%%%%%%%%%%%%%%%%%%%%%%%%%%%%%%%%%%%%%%
\section{Discussion}
In this paper we have present a gravity dual to the metal/insulator phase transition and found that the model can describe the CMR effect in some manganese oxides materials. The behavior of DC
resistivity is in complete agreement with experimental data.  The model provides a new example to apply the AdS/CFT correspondence to condensed matter systems.
As the first attempt to describe the CMR effect in a holographic setup,  more aspects of the model should be further studied. The first one is to study the behavior of DC resistivity in the case with
arbitrary graviton mass. For this, we need to consider the perturbations of the gravitational background, which is under investigating. Note that the current model  only realizes the macroscopic phenomenon in large scale. It is very interesting to consider whether one can directly realize such local electronic phase separation in a holographic setup  with Einstein's gravity theory rather than massive gravity. For example, we can set the chemical potential periodic in the spatial directions and take the lattice and impurity  into account. In those cases, we can expect that $\psi$ and $\rho$ are both inhomogeneous, electronic phase separation may be realized. For this, we have to deal with a set of partially differential equations and the involved numerical computation is extremely non-trivial. We expect it could be reported in future.

%%%%%%%%%%%%%%%%%%%%%%%%%%%%%%%%%%%%%%%%%%%%%%%%
\section*{Acknowledgements}
We thank helpful discussions with J. Erdmenger, S. Hartnool, F. Kunsmartsev, K. Schalm, J. Zaanen and many participants of the KITPC program ``Holographic duality for condensed matter physics" (July 6-31,2015, Beijing). This work was supported in part by the National Natural Science Foundation of China (No.11375247, and No.11435006 ).

%%%%%%%%%%%%%%%%%%%%%%%%%%%%%%%%%%%%%%%%%%%%%%%%%%%

%%%%%%%%%%%%%%%%%%%%%%%%%%%%%%%%


\begin{thebibliography}{99}
%\cite{Maldacena:1997re}
\bibitem{Maldacena:1997re}
  J.~M.~Maldacena,
  ``The large N limit of superconformal field theories and supergravity,''
  Adv.\ Theor.\ Math.\ Phys.\  {\bf 2}, 231 (1998)
  [Int.\ J.\ Theor.\ Phys.\  {\bf 38}, 1113 (1999)]
  [arXiv:hep-th/9711200].
  %%CITATION = IJTPB,38,1113;%%

%\cite{Gubser:1998bc}
\bibitem{Gubser:1998bc}
  S.~S.~Gubser, I.~R.~Klebanov and A.~M.~Polyakov,
  ``Gauge theory correlators from non-critical string theory,''
  Phys.\ Lett.\  B {\bf 428}, 105 (1998) [arXiv:hep-th/9802109].
  %%CITATION = PHLTA,B428,105;%%



%\cite{Witten:1998qj}
\bibitem{Witten:1998qj}
  E.~Witten,
  ``Anti-de Sitter space and holography,''
  Adv.\ Theor.\ Math.\ Phys.\  {\bf 2}, 253 (1998)
  [arXiv:hep-th/9802150].
  %%CITATION = 00203,2,253;%%

%\cite{Witten:1998qj2}
\bibitem{Witten:1998qj2}
E. Witten,
¡°Anti-de Sitter space, thermal phase transition, and confinement in gauge theories,¡±
Adv. Theor. Math. Phys. {\bf 2}, 505 (1998)
[arXiv:hep-th/9803131].


%\cite{Gubser:2008px}
\bibitem{Gubser:2008px}
  S.~S.~Gubser,
  %``Breaking an Abelian gauge symmetry near a black hole horizon,''
  Phys.\ Rev.\ D {\bf 78}, 065034 (2008)
  [arXiv:0801.2977 [hep-th]].
  %%CITATION = ARXIV:0801.2977;%%
  %547 citations counted in INSPIRE as of 05 juil. 2015


%\cite{Hartnoll:2008vx}
\bibitem{Hartnoll:2008vx}
  S.~A.~Hartnoll, C.~P.~Herzog and G.~T.~Horowitz,
  ``Building a Holographic Superconductor,''
  Phys.\ Rev.\ Lett.\  {\bf 101}, 031601 (2008)
  [arXiv:0803.3295 [hep-th]].
  %%CITATION = ARXIV:0803.3295;%%
  %527 citations counted in INSPIRE as of 05 Jun 2013


 %\cite{Lee:2008xf}
\bibitem{Lee:2008xf}
  S.~S.~Lee,
  ``A Non-Fermi Liquid from a Charged Black Hole: A Critical Fermi Ball,''
  Phys.\ Rev.\ D {\bf 79}, 086006 (2009)
  [arXiv:0809.3402 [hep-th]].
  %%CITATION = ARXIV:0809.3402;%%
  %214 citations counted in INSPIRE as of 11 Dec 2014

%\cite{Liu:2009dm}
\bibitem{Liu:2009dm}
  H.~Liu, J.~McGreevy and D.~Vegh,
  ``Non-Fermi liquids from holography,''
  Phys.\ Rev.\ D {\bf 83}, 065029 (2011)
  [arXiv:0903.2477 [hep-th]]
  %%CITATION = ARXIV:0903.2477;%%
  %237 citations counted in INSPIRE as of 05 Jun 2013

 %\cite{Cubrovic:2009ye}
\bibitem{Cubrovic:2009ye}
  M.~Cubrovic, J.~Zaanen and K.~Schalm,
  ``String Theory, Quantum Phase Transitions and the Emergent Fermi-Liquid,''
  Science {\bf 325}, 439 (2009)
  [arXiv:0904.1993 [hep-th]].
  %%CITATION = ARXIV:0904.1993;%%
  %253 citations counted in INSPIRE as of 11 Dec 2014


  %\cite{Cai:2015cya}
\bibitem{Cai:2015cya}
  R.~G.~Cai, L.~Li, L.~F.~Li and R.~Q.~Yang,
  %``Introduction to Holographic Superconductor Models,''
  Sci.\ China Phys.\ Mech.\ Astron.\  {\bf 58}, no. 6, 060401 (2015)
  [arXiv:1502.00437 [hep-th]].
  %%CITATION = ARXIV:1502.00437;%%
  %12 citations counted in INSPIRE as of 05 juil. 2015

%\cite{Cai:2014oca}
\bibitem{Cai:2014oca}
  R.~-G.~Cai and R.~-Q.~Yang,
  ``Paramagnetism-Ferromagnetism Phase Transition in a Dyonic Black Hole,''
  Phys. Rev. D {\bf 90}, 081901 (2014)
  [arXiv:1404.2856 [hep-th]].  %%CITATION = ARXIV:1404.2856;%%  %1 citations counted in INSPIRE as of 12 May 2014


 %\cite{Cai:2014jta}
\bibitem{Cai:2014jta}
  R.~G.~Cai and R.~Q.~Yang,
  ``Holographic model for the paramagnetism/antiferromagnetism phase transition,''
  Phys.\ Rev.\ D {\bf 91}, no. 8, 086001 (2015)
  [arXiv:1404.7737 [hep-th]].
  %%CITATION = ARXIV:1404.7737;%%
  %8 citations counted in INSPIRE as of 05 Jul 2015



 %\cite{Cai:2014dza}
\bibitem{Cai:2014dza}
  R.~G.~Cai and R.~Q.~Yang,
  ``Coexistence and competition of ferromagnetism and $p$-wave superconductivity in holographic model,''
  Phys.\ Rev.\ D {\bf 91}, no. 2, 026001 (2015)
  [arXiv:1410.5080 [hep-th]].
  %%CITATION = ARXIV:1410.5080;%%
  %5 citations counted in INSPIRE as of 05 Jul 2015

 %\cite{Cai:2015bsa}
\bibitem{Cai:2015bsa}
  R.~G.~Cai and R.~Q.~Yang,
  ``Antisymmetric tensor field and spontaneous magnetization in holographic duality,''
  [arXiv:1504.00855 [hep-th]].
  %%CITATION = ARXIV:1504.00855;%%
  %1 citations counted in INSPIRE as of 24 May 2015

 %\cite{Cai:2015mja}
\bibitem{Cai:2015mja}
  R.~G.~Cai, R.~Q.~Yang and F.~V.~Kusmartsev,
  ``A holographic model for antiferromagnetic quantum phase transition induced by magnetic field,''
  arXiv:1501.04481 [hep-th].
  %%CITATION = ARXIV:1501.04481;%%
  %3 citations counted in INSPIRE as of 05 Jul 2015

 %\cite{Cai:2015xpa}
\bibitem{Cai:2015xpa}
  R.~G.~Cai, R.~Q.~Yang and F.~V.~Kusmartsev,
  ``Holographic antiferromganetic quantum criticality and AdS$_2$ scaling limit,''
  arXiv:1505.03405 [hep-th].
  %%CITATION = ARXIV:1505.03405;%%

 %\cite{Cai:2015jta}
\bibitem{Cai:2015jta}
  R.~G.~Cai, R.~Q.~Yang, Y.~B.~Wu and C.~Y.~Zhang,
  ``Massive $2$-form field and holographic ferromagnetic phase transition,''
  arXiv:1507.00546 [hep-th].
  %%CITATION = ARXIV:1507.00546;%%


%\cite{Urushibara}
\bibitem{Urushibara}
A. Urushibara, Y. Moritomo, T. Arima, A. Asamitsu, G. Kido, Y. Tokura, ``Insulator-metal transition and giant magnetoresistance in La$_{1-x}$Sr$_x$MnO$_3$", Phys. Rev. B {\bf 51}, 14,103 (1995).

%\cite{Uehara}
\bibitem{Uehara}
Uehara, M., Mori, S., Chen, C. H., and Cheong, S. W., ``Percolative phase separation underlies colossal magnetoresistance in mixed-valent manganites", Nature, {\bf 399} (6736), 560-563 (1999).

%\cite{MBS}
\bibitem{MBS}
M. B. Salamon, and M. Jaime, ``The physics of manganites: Structure and transport",  Rev. Mod. Phys. {\bf 73}, 583 (2001).

%\cite{Dagottoa}
\bibitem{Dagottoa}
E. Dagotto, T. Hotta, A. Moreo, ``Colossal magnetoresistant materials: the key role of phase separation", Phy. Rep. {\bf 344}, 1-153 (2001).

%\cite{Nagaev}
\bibitem{Nagaev}
E.L. Nagaev, ``Colossal-magnetoresistance materials: manganites and conventional ferromagnetic semiconductors", Phys. Rep. {\bf 346}, 387-531 (2001).

%\cite{Cengiz}
\bibitem{Cengiz}
Cengiz Sen, Gonzalo Alvarez, Elbio Dagotto, ``Unveiling First Order CMR Transitions in the Two-Orbital Model for Manganites", Phys. Rev. Lett. {\bf 105}, 097203 (2010).

%\cite{Mukherjee}
\bibitem{Mukherjee}
Mukherjee Anamitra, Cole William S., Woodward Patrick, Randeria Mohit and Trivedi Nandini, ``Theory of Strain-Controlled Magnetotransport and Stabilization of the Ferromagnetic Insulating Phase in Manganite Thin Films", Phys. Rev. Lett. {\bf 110}, 157201 (2013).

%\cite{Dagottoa2}
\bibitem{Dagottoa2}
E. Dagotto ``Open questions in CMR manganites, relevance of clustered states and analogies with other compounds including the cuprates",
 New J. Phys. {\bf 7}, 67 (2005) [arXiv:cond-mat/0302550].

%\cite{Blake:2013owa}
\bibitem{Blake:2013owa}
  M.~Blake, D.~Tong and D.~Vegh,
  ``Holographic Lattices Give the Graviton an Effective Mass,''
  Phys.\ Rev.\ Lett.\  {\bf 112}, no. 7, 071602 (2014)
  [arXiv:1310.3832 [hep-th]].
  %%CITATION = ARXIV:1310.3832;%%
  %58 citations counted in INSPIRE as of 17 Jul 2015

%\cite{Mefford:2014gia}
\bibitem{Mefford:2014gia}
  E.~Mefford and G.~T.~Horowitz,
  ``Simple holographic insulator,''
  Phys.\ Rev.\ D {\bf 90}, no. 8, 084042 (2014)
  [arXiv:1406.4188 [hep-th]].
  %%CITATION = ARXIV:1406.4188;%%
  %10 citations counted in INSPIRE as of 05 Jul 2015



%\cite{deRham:2010ik}
\bibitem{deRham:2010ik}
  C.~de Rham and G.~Gabadadze,
  ``Generalization of the Fierz-Pauli Action,''
  Phys.\ Rev.\ D {\bf 82}, 044020 (2010)
  [arXiv:1007.0443 [hep-th]].
  %%CITATION = ARXIV:1007.0443;%%
  %482 citations counted in INSPIRE as of 17 juil. 2015

%\cite{deRham:2010kj}
\bibitem{deRham:2010kj}
  C.~de Rham, G.~Gabadadze and A.~J.~Tolley,
  ``Resummation of Massive Gravity,''
  Phys.\ Rev.\ Lett.\  {\bf 106}, 231101 (2011)
  [arXiv:1011.1232 [hep-th]].
  %%CITATION = ARXIV:1011.1232;%%
  %593 citations counted in INSPIRE as of 17 Jul 2015

%\cite{Hassan:2011vm}
\bibitem{Hassan:2011vm}
  S.~F.~Hassan and R.~A.~Rosen,
  ``On Non-Linear Actions for Massive Gravity,''
  JHEP {\bf 1107}, 009 (2011)
  [arXiv:1103.6055 [hep-th]].
  %%CITATION = ARXIV:1103.6055;%%
  %189 citations counted in INSPIRE as of 17 Jul 2015

%\cite{Blake:2013bqa}
\bibitem{Blake:2013bqa}
  M.~Blake and D.~Tong,
  ``Universal Resistivity from Holographic Massive Gravity,''
  Phys.\ Rev.\ D {\bf 88}, no. 10, 106004 (2013)
  [arXiv:1308.4970 [hep-th]].
  %%CITATION = ARXIV:1308.4970;%%
  %58 citations counted in INSPIRE as of 18 juil. 2015

%\cite{Wiltshire:1994de}
\bibitem{Wiltshire:1994de}
  D.~L.~Wiltshire,
  ``Dilaton black holes with a cosmological term,''
  J.\ Austral.\ Math.\ Soc.\ B {\bf 41}, 198 (1999)
  [gr-qc/9502038].
  %%CITATION = GR-QC/9502038;%%
  %10 citations counted in INSPIRE as of 28 May 2015

%\cite{Davison:2013jba}
\bibitem{Davison:2013jba}
  R.~A.~Davison,
  ``Momentum relaxation in holographic massive gravity,''
  Phys.\ Rev.\ D {\bf 88}, 086003 (2013)
  [arXiv:1306.5792 [hep-th]].
  %%CITATION = ARXIV:1306.5792;%%
  %57 citations counted in INSPIRE as of 17 Jul 2015

%\cite{Vegh:2013sk}
\bibitem{Vegh:2013sk}
  D.~Vegh,
  ``Holography without translational symmetry,''
  arXiv:1301.0537 [hep-th].
  %%CITATION = ARXIV:1301.0537;%%
  %77 citations counted in INSPIRE as of 17 Jul 2015


%\cite{Iqbal:2008by}
\bibitem{Iqbal:2008by}
  N.~Iqbal and H.~Liu,
  ``Universality of the hydrodynamic limit in AdS/CFT and the membrane paradigm,''
  Phys.\ Rev.\ D {\bf 79}, 025023 (2009)
  [arXiv:0809.3808 [hep-th]].
  %%CITATION = ARXIV:0809.3808;%%
  %272 citations counted in INSPIRE as of 25 Dec 2014

%\cite{P.Schiffer}
%\bibitem{P.Schiffer}
%P. Schiffer, A. P. Ramirez, W. Bao, and S-W. Cheong, ``Low Temperature Magnetoresistance and the Magnetic Phase Diagram of La$_{1-x}$Ca$_x$MnO$_3$", %Phys. Rev. Lett. {\bf 75}, 3336 (1995).



%\cite{Eric}
%\bibitem{Eric}
%Eric Mefford and Gary T. Horowitz, ``Simple holographic insulator", Phys. Rev. D {\bf 90}, 084042 (2014).


%\cite{TFau}
%\bibitem{TFau}
%T. Faulkner, G. T. Horowitz, and M. M. Roberts, ``Holographic quantum criticality from multi-trace deformations", JHEP {\bf 04} (2011) 051.



%\cite{SW.Cheonga}
\bibitem{SW.Cheonga}
S.-W. Cheong, P.A. Sharma, N. Hur, Y. Horibe and C.H. Chen, ``Electronic phase separation in complex materials", Physica B {\bf 318} (2002) 39-51.


\end{thebibliography}
\end{document}